\documentclass[aps,pre,twocolumn,groupedaddress,showpacs,showkeys]{revtex4}

\usepackage{graphicx}
\usepackage{dcolumn}
\usepackage{bm}
\usepackage{amssymb}
\usepackage{amsmath}
\usepackage[usenames]{color}

\newcommand{\lulu}[1]{#1}

\begin{document}


\title{\lulu{Generalized synchronization in mutually coupled oscillators and complex networks}}


\author{Olga~I.~Moskalenko$^{1}$}
\email{o.i.moskalenko@gmail.com}
\author{Alexey~A.~Koronovskii$^{1}$}
\email{alkor@nonlin.sgu.ru}
\author{Alexander~E.~Hramov$^{1}$}
\email{aeh@nonlin.sgu.ru}
\author{Stefano Boccaletti$^{2}$}
\affiliation{$^1$ Faculty of Nonlinear Processes, Saratov State
University, Astrakhanskaya, 83, Saratov, 410012, Russia \\
Saratov State Technical University, Politehnicheskaya, 77, Saratov,
410054, Russia \\$^2$ CNR --- Istituto dei Sistemi Complessi Via
Madonna del Piano, 10 50019 Sesto Fiorentino (FI), Italy}



\date{\today}

\begin{abstract}
We introduce a novel concept of generalized synchronization, able to
encompass the setting of collective synchronized behavior for
mutually coupled systems and networking systems featuring complex
topologies in their connections. The onset of the synchronous regime
is confirmed by the dependence of the system's Lyapunov exponents on
the coupling parameter. The presence of a generalized
synchronization regime is verified by means of the nearest neighbor
method.
\end{abstract}


\maketitle


\section*{INTRODUCTION}
\label{sct:Introduction}

Synchronization of chaotic systems is a subject widely studied in
recent years, having both theoretical and applied
significance~\cite{Boccaletti:2002_ChaosSynchro}. Of a particular
interest is the intricate phenomenon of generalized synchronization
(GS), originally introduced as an emerging collective motion of
unidirectionally coupled chaotic
oscillators~\cite{Rulkov:1996_AuxiliarySystem,
Aeh:2005_GS:ModifiedSystem}.
GS has been, indeed, observed in numerous systems, both
numerically~\cite{Kocarev:1996_GS,Zhigang:2000_GSversusPS,Hramov:2005_GLEsPRE}
and
experimentally~\cite{Rulkov:1996_SynchroCircuits,GS_LightModulator,dmitriev:074101},
and its main
features~\cite{Zhigang:2000_GSversusPS,Hramov:2008_INIS_PRE} have
suggested many possible
applications~\cite{Terry:GSchaosCom2001,alkor:2010_SecureCommunicationUFNeng,Moskalenko:InfoTransNoisePLA2010}.

The very same concept of GS has been introduced initially in the case of two
unidirectionally coupled oscillators, the drive (or master)
$\mathbf{x}(t)$ system and the response (or slave) $\mathbf{u}(t)$ one. GS means the
presence of a time independent functional relation $\mathbf{F}[\cdot]$
between the master and slave system states, after a suitable transient time interval is
elapsed~\cite{Rulkov:1995_GeneralSynchro}, i.e.
\begin{equation}\label{eq:FunctRel}
{\mathbf{u}(t)=\mathbf{F}[\mathbf{x}(t)]}.
\end{equation}

In this Paper, we report on the extension of the concept of GS for
both oscillators with a bidirectional coupling, and for complex
networks' architectures.
It is, indeed, evident that Eq.~(\ref{eq:FunctRel}) needs to be extended for
reflecting the fact that a mutual interaction exists between
the systems under study.
The traditional definition of GS in the form of~(\ref{eq:FunctRel}) introduced for the unidirectionally coupled chaotic oscillators is based on the fact that the drive system state $\mathbf{x}(t)$ does not depend on the state of the response system $\mathbf{u}(t)$. Hence, for the GS regime Eq.~(\ref{eq:FunctRel}) may be applied at every moment of time $t$. In the case of the mutual type of coupling between systems the situation is radically different, since the interacting oscillators are equivalent from the point of view of the coupling, and the evolution of the first system is determined not only by the vector $\mathbf{x}(t)$, but also by the the state of the second system $\mathbf{u}(t)$, and vice versa. Therefore, in the general case the functional relation between states of the mutually coupled systems can not be written in the explicit form~(\ref{eq:FunctRel}), and, as a consequence, the implicit functional relation must be used. To take into account the common action of
the systems on each other, we here propose to modify Eq.~(\ref{eq:FunctRel})
for two chaotic oscillators as
\begin{equation}
\mathbf{F}[\mathbf{x}(t),\mathbf{u}(t)]=0. \label{eq:MutualFunctRel}
\end{equation}
Note, Eq.~(\ref{eq:MutualFunctRel}) may be used to inspect both the unidirectional and the bidirectional types
of the coupling (as well as any coupling with arbitrary asymmetry between the systems). Furthermore, Eq.~(\ref{eq:FunctRel}) can be considered
as a special case of (\ref{eq:MutualFunctRel}).

For a generic ensemble of $N$ elements
Eq.~(\ref{eq:NetworkFunctRel}) can be rewritten as following
\begin{equation}
\mathbf{F}[\mathbf{x}_1(t),\mathbf{x}_2(t),\dots,\mathbf{x}_i(t),\dots,\mathbf{x}_N(t)]=0, \label{eq:NetworkFunctRel}
\end{equation}
where $\mathbf{x}_i(t)$ is the vector state of the $i$-th element of
the network. In other words, we assume that the generalized
synchronization regime implies the presence of a functional relation
between the system states, as before, but this relation takes the
implicit form~(\ref{eq:NetworkFunctRel}) instead of
(\ref{eq:FunctRel}), i.e. it implies the arousal of a generic
manifold in phase-space wherein the overall trajectory is lying
during its time evolution.

\section{Generalized synchronization, Lyapunov exponents and the nearest neighbor method}
\label{sct:LE&NNM}

For the unidirectionally coupled oscillators the Lyapunov exponents calculation is known to allow to detect the GS boundary more precisely in comparison with the nearest neighbor method, whereas the last one gives mostly only the qualitative confirmation of the GS regime presence. Therefore, we propose to use the Lyapunov exponents to detect the GS regime in the mutually coupled systems. As far as the nearest neighbor method is concerned, we use it for the verification of the obtained results.

If the dimensions of the drive and response
systems are $N_d$ and $N_r$ respectively, the behavior of the
unidirectionally coupled oscillators is
characterized by the Lyapunov exponent spectrum
${\lambda_1\geq\lambda_2\geq\dots\geq\lambda_{N_d+N_r}}$. Due to the
independence of the drive system dynamics on the behavior of the
response one, the Lyapunov exponent spectrum may be divided into two
parts~\cite{Pecora:1991_ChaosSynchro,Pyragas:1997_CLEsFromTimeSeries}:
LEs of the drive system ${\lambda^d_1\geq\dots\geq\lambda^d_{N_d}}$
and LEs of the response one
${\lambda^r_1\geq\dots\geq\lambda^r_{N_r}}$. For the GS detection in
the unidirectionally coupled chaotic oscillators the Lyapunov
exponents calculated for the response system play the key role
(since the behavior of the response depends on the drive, these
Lyapunov exponents are called conditional). With the increase of the
coupling strength the largest conditional Lyapunov exponent
${\lambda^r_1}$ becomes negative at the onset of the GS
regime~\cite{Pyragas:1996_WeakAndStrongSynchro}. Thus, the
negativity of the largest conditional Lyapunov exponent
\begin{equation}\label{eq:CLEGSCondition}
\lambda^r_1<0
\end{equation}
is considered as a criterion of the GS presence in
the unidirectionally
coupled dynamical systems~\cite{Pyragas:1997_CLEsFromTimeSeries,%
Aeh:2005_GS:ModifiedSystem}.

Since in the case of the mutual coupling the spectrum of Lyapunov exponents cannot be divided into two parts corresponding to the drive and response systems, condition~(\ref{eq:CLEGSCondition}) must also be modified. Notice that, here, the dimension of each element is assumed to be $N_d=3$, but this analytical study may be
extended easily to the other systems with arbitrary dimensions $N_d$.

As it has already been mentioned, in the neighborhood of any
moment of time $t$~\footnote{Except, may be, the finite number of
points, which may be eliminated from the consideration without the
lack of generality} Eq.~(\ref{eq:MutualFunctRel}) may be considered
as the definition of the implicit functional relation between system
states, and, therefore, according to the implicit-function theorem~\cite{Springer:2002_EncyclopaediaofMathematics}, locally,
the implicit definition of the functional relation between system
states may be used, i.e., $\mathbf{x}(t)=\mathbf{\tilde{F}}[\mathbf{u}(t)]$ or
$\mathbf{u}(t)=\mathbf{\tilde{F}}[\mathbf{x}(t)]$.
Let us assume without the lack of the
generality that for $t^*-\delta<t<t^*+\delta$ (where $\delta$ is
infinitely small) the implicit functional relation
\begin{equation}\label{eq:FuncRelReduced}
\mathbf{x}(t^*)=\mathbf{\tilde{F}}[\mathbf{u}(t^*)]
\end{equation}
is defined with the help of Eq.~(\ref{eq:MutualFunctRel}). In this case, locally, in the given range $t^*-\delta<t<t^*+\delta$, we deal with the already studied case of Eq.~(\ref{eq:FunctRel}). Eq.~(\ref{eq:FuncRelReduced}) means that (under assumption $N_d=3$ made above) the following local Lyapunov exponents characterize the dynamics of the systems: $\lambda^u_1>0$, $\lambda^u_2=0$, $\lambda^u_3<0$, $\lambda^x_{1,2,3}<0$. In other words, in the selected area of the $6D$ phase-space the manifold corresponding to the GS regime is characterized by one unstable direction $\mathbf{e}^u$ and one direction with the neutral stability $\mathbf{e}^0$ lying inside this manifold, whereas all other directions are stable. These directions correspond to one positive, one zero and four negative Lyapunov exponents. The same statement is also correct for the other moments of time $t^*$, although the implicit functional relation may take form $\mathbf{u}(t^*)=\mathbf{\tilde{F}}[\mathbf{x}(t^*)]$ instead of (\ref{eq:FuncRelReduced}). So, one can come to conclusion that in the case of the mutual type of coupling the manifold corresponding to the GS regime at every moment of time is characterized by one unstable direction, one direction with the neutral stability and four stable directions. So, having calculated the spectrum of Lyapunov exponents for two bidirectionally coupled chaotic oscillators one obtain that the GS regime in this case produces one positive, one zero and four negative Lyapunov exponents. Since in the case of the mutual coupling one cannot pick out the conditional Lyapunov exponents, the criterion of the GS regime should be written in the form
\begin{equation}\label{eq:BiDirLEGSCondition}
\lambda_3<0,
\end{equation}
whereas $\lambda_1>0$, $\lambda_2=0$.

Having extended this
theoretical consideration to the complex networks (where the
implicit form of the functional relation is given by
(\ref{eq:NetworkFunctRel})) one can obtain that
condition~(\ref{eq:BiDirLEGSCondition}) remains also correct for the
GS regime existence in these networks.

To validate the presence of the generalized synchronization
regime, in parallel with the calculation of  the spectrum of
Lyapunov exponents, one can also make use of the nearest neighbor
method~\cite{Rulkov:1995_GeneralSynchro,Parlitz:1996_PhaseSynchroExperimental}.
The main idea of this technique consists in the fact that the
presence of the functional relation between the interacting system
states means that all close states (``origins'') in the phase
sub-space of a given system $\mathbf{x}(t)$ should correspond to
close states (``images'') in the phase sub-space of the other
one $\mathbf{u}(t)$ (see \cite{Rulkov:1995_GeneralSynchro}
for details). For mutually coupled oscillators the inverse statement
must be also correct, i.e. all close states in the phase sub-space
of the second system $\mathbf{u}(t)$ must correspond to close states
of the first one $\mathbf{x}(t)$.

As a numerical indicator of the existence of a functional
relationship between the interacting systems, the mean distance
between images $\mathbf{u}^{k,kn}$ of nearest neighbors
$\mathbf{x}^{k,kn}$ normalized by the average distance $\delta$ of
randomly chosen states of the first system, i.e.
\begin{equation}
d=\frac{1}{M\delta}\sum_{k=0}^{M-1}||\mathbf{u}^k-\mathbf{u}^{kn}||,
\label{eq:ClosenessMeasure}
\end{equation}
can be calculated (here $M$ is the number of the points chosen
randomly)~\cite{Parlitz:1996_PhaseSynchroExperimental}. This
characteristic allows to reveal the qualitative changes in the
synchonous/asynchonous behavior of the coupled systems. When the
coupling between systems is very small and oscillators show the
asynchronous dynamics the value of this measure is $d\sim 1$. The
phase synchronization regime causes the sharp decrease of the value
$d$ but it differs from zero sufficiently, as before, whereas deep
inside the generalized synchronization region $d$ tends to be zero
due to the presence of the functional relation between states of the
interacting systems. In the vicinity of the GS boundary (both inside
and outside the GS area) the value of $d$-quantity decreases slowly
to zero value, with the onset of the generalized synchronization
regime corresponding approximately to the middle of this transition
interval. Unfortunately, the nearest neighbor method does not allow
to detect precisely the boundary points of the GS regime, but it
allows to confirm the presence of GS and to estimate the location of
the GS boundary.

To illustrate the proposed approach, we use several cases, and start
our considerations from two mutually coupled R\"ossler
oscillators

\begin{equation}
\begin{array}{ll}
\dot x_{1,2}=-\omega_{1,2}y_{1,2}-z_{1,2} +\varepsilon(x_{2,1}-x_{1,2}),\\
\dot y_{1,2}=\omega_{1,2}x_{1,2}+ay_{1,2},\\
\dot z_{1,2}=p+z_{1,2}(x_{1,2}-c),
\end{array}
\label{eq:Roesslers}
\end{equation}
where $\mathbf{x}_{1,2}(t)=(x_{1,2},y_{1,2},z_{1,2})^T$ are the
vector-states of the interacting systems, $\varepsilon$ is a
coupling parameter. The control parameter values have been selected
by analogy with our previous
works~\cite{Harmov:2005_GSOnset_EPL,Hramov:2007_2TypesPSDestruction,Hramov:ZeroLE_PRE2008}
as $a=0.15$, $p=0.2$, $c=10$. The parameter $\omega_{1,2}$ defines
the natural frequency of oscillations, with $\omega_1=0.99$ and $\omega_2=0.95$ being
fixed.

\begin{figure}[t]
\includegraphics[width=8.7cm]{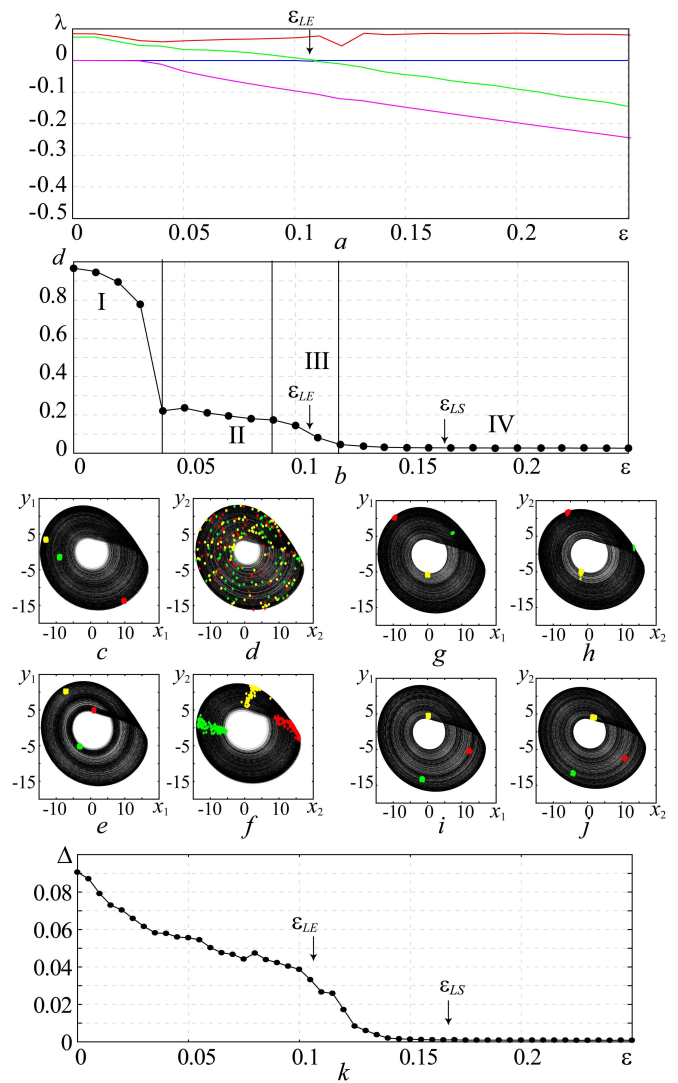}
\caption{(Color online) (\textit{a}) The dependencies of the four largest
Lyapunov exponents on the coupling strength $\varepsilon$ for (\ref{eq:Roesslers}). (\textit{b}) The quantitative measure $d$
(\ref{eq:ClosenessMeasure}) versus the coupling parameter strength
$\varepsilon$. The critical values of the coupling parameter
$\varepsilon_{LE}=0.106$ and
$\varepsilon_{LS}=0.169$ are marked
by arrows. (\textit{c--j}) The phase portraits of R\"ossler
oscillators for the different values of the coupling parameter:
(\textit{c-d}) $\varepsilon=0.01$ (the asynchronous state);
(\textit{e-f}) $\varepsilon=0.05$ (the PS regime); (\textit{g-h})
$\varepsilon=0.12$ (the GS regime); (\textit{i-j})
$\varepsilon=0.18$ (the LS regime).
Figures (\textit{c,e,g,i}) show the chaotic attractor of the first system $\mathbf{x}(t)$ with three randomly chosen points $\mathbf{x}^{k}$
and its nearest neighbors $\mathbf{x}^{kn}$. Figures~(\textit{d,f,h,j}) illustrate the corresponding
states $\mathbf{u}^{k,kn}$ in the phase sub-space of the second system
$\mathbf{u}(t)$. (\textit{k}) The dependence of the mean distance $\Delta$ between shifted states of the interacted systems~(\ref{eq:Roesslers})}
\label{fgr:TwoRoesslers}
\end{figure}

The behavior of the four largest Lyapunov exponents for the considered case is shown in Fig.~\ref{fgr:TwoRoesslers},\,\textit{a} and the dependence of the quantitative measure
(\ref{eq:ClosenessMeasure}) on the coupling parameter strength
$\varepsilon$ is given in Fig.~\ref{fgr:TwoRoesslers},\textit{b}. One can see that at
$\varepsilon_{LE}\approx0.106$ the second Lyapunov exponent
$\lambda_2$ passes through zero and becomes negative. Therefore, the
generalized synchronization regime is expected to be observed above
the critical value $\varepsilon_{LE}$. As far as the quantitative measure
(\ref{eq:ClosenessMeasure}) is concerned, the curve $d(\varepsilon)$ decreases monotonically when the coupling
parameter value increases. The $(\varepsilon;d)$-plane can be
divided into 4 parts: I -- ${\varepsilon\in[0;0.04)}$, the
$d$--characteristic decreases very sharply indicating the transition
from the asynchronous motion to the phase synchronization regime at
${\varepsilon_{PS}=0.04}$; II -- ${\varepsilon\in[0.04;0.09)}$, $d$ is not practically changed being the evidence of the phase synchronized motion; III -- ${\varepsilon\in[0.09;0.12)}$, the
$d$--characteristic decreases slowly indicating the occurrence of
the GS regime; IV -- ${\varepsilon>0.12}$, ${d\approx 0}$.
Note, the bifurcation point $\varepsilon_{LE}\approx0.106$ (where the second Lyapunov exponent
$\lambda_2$ crosses the zero value and becomes negative) corresponds exactly to the middle of the transition interval III where $d$--characteristic decreases slowly to zero value. Later, at $\varepsilon_{LS}\approx 0.169$ the lag synchronization (LS) regime takes place.

In Fig.~\ref{fgr:TwoRoesslers},\textit{c-j} the phase portraits of
the interacting R\"ossler systems~(\ref{eq:Roesslers}) are shown for
the different values of the coupling strength $\varepsilon$. In the
phase portraits of the first system $\mathbf{x}(t)$
(Fig.~\ref{fgr:TwoRoesslers},\,\textit{c,e,g,i}) three randomly
chosen points $\mathbf{x}^{k}$ with its nearest neighbors
$\mathbf{x}^{kn}$ are shown.
Figures~\ref{fgr:TwoRoesslers},\textit{d,f,h,j} illustrate the
corresponding states $\mathbf{u}^{k,kn}$ in the phase sub-space of
the second system $\mathbf{u}(t)$.

One can see easily that for the small values of the coupling
parameter ($\varepsilon=0.01$) all points of the second system are
distributed randomly throughout the whole attractor
(Fig.~\ref{fgr:TwoRoesslers},\,\textit{d}). When the coupling parameter value
increases the points become to be concentrated in a limited portion of
attractor, with the radius of the distribution area decreasing
(compare Fig.~\ref{fgr:TwoRoesslers},\textit{f,h}). For
$\varepsilon>\varepsilon_{LE}$ all states of the second system
corresponding to the nearest states of the first oscillator are also
nearest, and vice versa (Fig.~\ref{fgr:TwoRoesslers},\,\textit{g,h} and Fig.~\ref{fgr:TwoRoesslers},\,\textit{i,j}), that proves the occurrence of GS.
Note also the difference between GS and LS which consists in the fact that in the LS regime the
representation points corresponding to the nearest neighbors are
practically in the same part of chaotic attractor (Fig.~\ref{fgr:TwoRoesslers},\,\textit{i,j})
whereas in the GS regime they can be located in the slightly
different regions (Fig.~\ref{fgr:TwoRoesslers},\,\textit{g,h}).
The additional evidence of the fact that the GS and LS regimes differ from each other is the behavior of the mean distance $\Delta$ between the interacted system states shifted in time on the coupling parameter value. Such dependence is shown in Fig.~\ref{fgr:TwoRoesslers},\textit{k}. It is clearly seen that in the LS regime $\Delta\approx 0$ whereas in the GS regime it is a positive one.

\lulu{\section{GS in mutually coupled spatially extended systems}
\label{sct:MutuallyCoupledPierceDiodes}}

The very same results have also been obtained for two mutually
coupled Pierce diodes being the classical models of beam-plasma
systems, demonstrating the complex spatio-temporal oscillations
including the chaotic ones~\cite{Godfrey:1987, Matsumoto:1996}. The
dynamics of Pierce diodes (in the fluid electronic approximation) is
described by the self-consistent system of dimensionless Poisson,
continuity and motion
equations~\cite{TrueAeh:2003_2004_microwave_electronics_1_2engl}:
\begin{equation}\label{eq:PierceDiodes}
\begin{array}{c}
\displaystyle
\frac{\partial^2\varphi_{1,2}}{\partial
x^2}=-\left(\alpha_{1,2}\right)^2(\rho_{1,2}-1),\\
\displaystyle
\frac{\partial \rho_{1,2}}{\partial t}=-\frac {\partial(\rho_{1,2}
v_{1,2})} {\partial x},\\
\displaystyle
\frac{\partial v_{1,2}}{\partial t}= -v_{1,2} \frac{\partial
v_{1,2}}{\partial x}+\frac{\partial \varphi_{1,2}}{\partial
x},
\end{array}
\end{equation}
with the boundary conditions
\begin{equation}\label{q4_ch8}
v_{1,2}(0,t)=1,\quad \rho_{1,2}(0,t)=1,\quad\varphi_{1,2}(0,t)=0,
\end{equation}
where $\varphi_{1,2}(x,t)$ is the dimensionless potential of the
electric field, $\rho_{1,2}(x,t)$ and $v_{1,2}(x,t)$ are the
dimensionless density and velocity of the electron beam ($0\leq x
\leq1$), the indexes ``1'' and ``2'' correspond to the first and
second coupled beam-plasma systems, respectively,
$\alpha_1=2.858\pi$, $\alpha_2=2.860\pi$ are the control parameters.
The bidirectional coupling between considered Pierce diodes is
realized by the modification of the boundary conditions on the right
boundary of the systems, in the same way as it has been done
in~\cite{Filatov:2006_PierceDiode_PLA,filatova:023133}
\begin{equation}
\varphi_{1,2}(1,t)=\varepsilon(\rho_{1,2}(x=1,t)-\rho_{2,1}(x=1,t)),
\label{eq:Coupling}
\end{equation}
where $\varepsilon$ is a dimensionless coupling parameter.

Continuity and motion equations of~(\ref{eq:PierceDiodes}) are
integrated numerically with the help of the one-step explicit
two-level scheme with upstream differences and the Poisson equation
is solved by the method of the error vector
propagation~\cite{Rouch:1976_FluidNumericalBook}. The time and space
integration steps have been taken as $\Delta t=0.003$ and $\Delta
x=0.005$, respectively.

\begin{figure}[p]
\includegraphics[width=8.2cm]{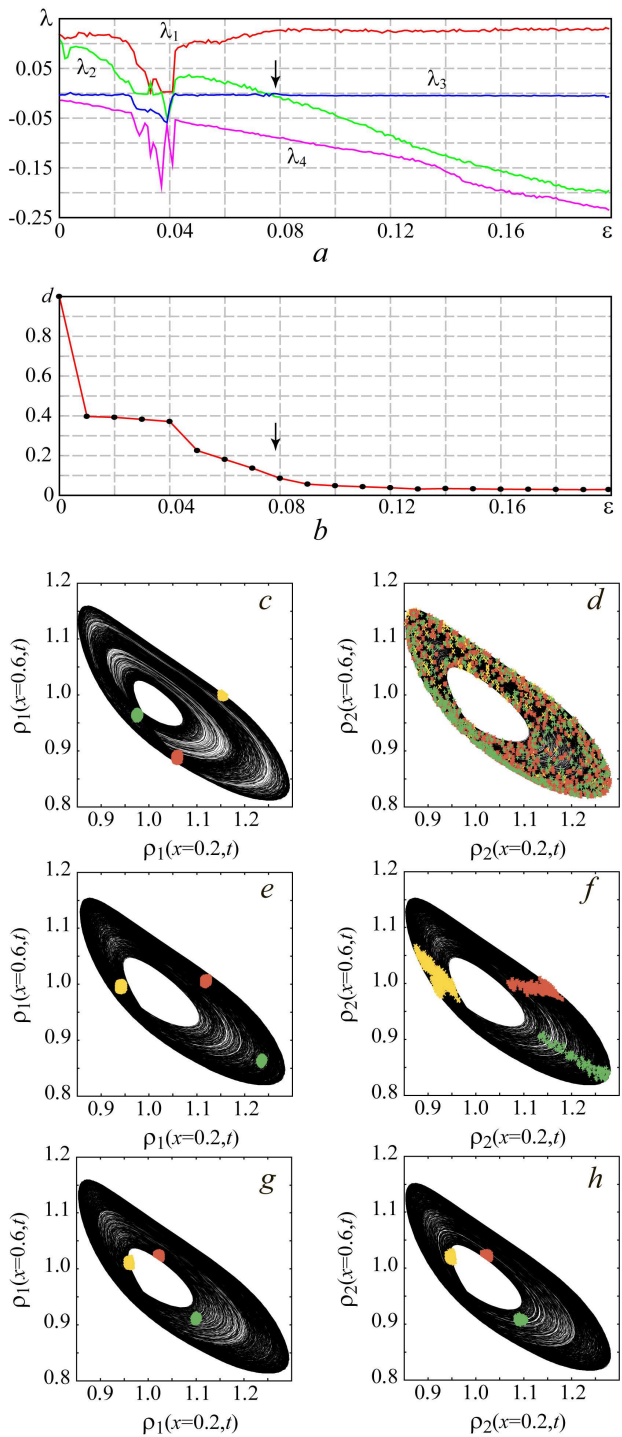}
\caption{(Color online) (\textit{a}) The dependencies of the four largest
Lyapunov exponents on the coupling strength $\varepsilon$ for (\ref{eq:PierceDiodes}). (\textit{b}) The quantitative measure $d$
(\ref{eq:ClosenessMeasure}) versus the coupling parameter strength
$\varepsilon$. The critical value of the coupling parameter
$\varepsilon_{LE}=0.078$ corresponding to the zero-cross of the positive Lyapunov exponent is marked
by arrow. (\textit{c--h}) The reconstructed attractors of two mutually coupled Pierce diodes on $(\rho_{1,2}(x=0.2,t),\rho_{1,2}(x=0.6,t))$-plane for
the different values of
the coupling parameter: (\textit{c,d}) $\varepsilon=0.002$ (the asynchronous state),
(\textit{e,f}) $\varepsilon=0.05$ (the PS regime),
(\textit{g,h}) $\varepsilon=0.10$ (the GS regime).
Figures (\textit{c,e,g}) show the reconstructed attractor of the first
Pierce diode with three randomly chosen points
and its nearest neighbors.
Figures~(\textit{d,f,h}) illustrate the corresponding states
in the phase sub-space of the second Pierce diode}
\label{fgr:LEPierceBidir}
\end{figure}

For the GS regime detection the nearest neighbor method and
calculation of the spectrum of Lyapunov exponents have also been
used. In Fig.~\ref{fgr:LEPierceBidir},\textit{a} the dependencies of
four largest Lyapunov exponents on the coupling parameter
$\varepsilon$ are shown. For computation of the spectrum of spatial
Lyapunov exponents the method proposed in~\cite{hramov:SLE_POP2012}
has been used. It is clearly seen that as in the case of mutually
coupled R\"ossler systems two Lyapunov exponents do not practically
depend on the coupling parameter $\varepsilon$, i.e. one Lyapunov
exponent $\lambda_1$ is always positive (except the windows of
periodicity) whereas the second one $\lambda_3$ remains zero. At the
same time, two Lyapunov exponents (initially positive $\lambda_2$
and initially zero $\lambda_4$) depend on the coupling parameter and
pass in the field of the negative values with the coupling parameter
value increasing. One can assume, that as in the case of mutually
coupled R\"ossler systems the transition of the positive Lyapunov
exponent $\lambda_2$ in the field of the negative values (for
$\varepsilon=\varepsilon_c=0.078$) is connected with the generalized
synchronization regime onset in mutually coupled beam-plasma
systems.

To confirm the assumption made above the nearest neighbor method has
been used. To characterize the degree of closeness of the interacted
system states quantitatively the measure $d$ defined by
(\ref{eq:ClosenessMeasure}) has also been computed. As the
interacted system states vectors
$\mathbf{u}_{1,2}(x,t)=(\rho_{1,2},v_{1,2},\varphi_{1,2})^T$ have
been used, whose norm $||\cdot||$ has been calculated as
\begin{equation}
||\mathbf{u}||=\sqrt{\int_0^L \rho\,dx+\int_0^L \varphi\,dx+\int_0^L v\,dx}.
\end{equation}
In Fig.~\ref{fgr:LEPierceBidir},\textit{b} the dependence
of the quantitative measure $d$ on the coupling parameter
$\varepsilon$ is shown. It is clearly seen that $d$-characteristics
decreases monotonically from one to zero with $\varepsilon$ value
increasing. At that, $\varepsilon_c$ is approximately in the middle
of the falling field $\varepsilon\in[0.04;0.12]$, that indicates the
generalized synchronization regime presence. It should be noted that
generalized synchronization does not coincide with the complete
synchronization regime in this case. For the control parameter
values mentioned above it is realized for $\varepsilon\approx 0.17$.

Additional evidence of the presence of generalized synchronization
regime in two mutually coupled beam plasma systems is the behavior of
the nearest neighbors in the phase space of interacted systems. In
Fig.~\ref{fgr:LEPierceBidir},\textit{c-h} the reconstructed
attractors of interacted Pierce diodes on
$(\rho_{1,2}(x=0.2,t),\rho_{1,2}(x=0.6,t))$-plane for different
values of the coupling parameter $\varepsilon$ are shown. On
attractors of the first system
(Fig.~\ref{fgr:LEPierceBidir},\textit{c,e,g}) three randomly chosen
points and its nearest neighbors are also indicated.
Fig.~\ref{fgr:LEPierceBidir},\textit{d,f,h} illustrates the
corresponding states in the phase space of the second system.

It is clearly seen that as in the case of R\"ossler systems for
small values of the coupling parameter ($\varepsilon=0.002$) all
points in the phase space of the second system are distributed
randomly over all attractor (see
Fig.~\ref{fgr:LEPierceBidir},\,\textit{d}). With the coupling
parameter value increasing the points begin grouping in the limited
range of attractor that corresponds to the phase synchronization
regime onset, with the radius of such field being decreased (compare
Fig.~\ref{fgr:LEPierceBidir},\,\textit{f,h}). For
$\varepsilon>\varepsilon_{LE}$ all states of the second beam-plasma
system corresponding to the nearest neighbors of the first Pierce
diode are also nearest and vise versa
(Fig.~\ref{fgr:LEPierceBidir},\,\textit{g,h}), that is the evidence
of the generalized synchronization regime presence.

So, on the basis of consideration carried out one can conclude that
in both cases considered above, the onset of the GS regime is
connected with the sign change of the initially positive Lyapunov
exponent $\lambda_2$ in the same way as for two unidirectional
oscillators~\cite{Pyragas:1996_WeakAndStrongSynchro}. As well as in
the case of the chaotic oscillators coupled unidirectionally, for
the systems with the bidirectional coupling the onset of the GS
regime $\varepsilon_{LE}$ precedes the boundary of lag or complete
synchronization $\varepsilon_{LS}$.

\section{GS in mutually coupled Lorenz oscillators}
\label{sct:MutuallyCoupledLrnzs}

To prove the made decisions conclusively, in this Section of the
paper we  consider another example of the oscillators coupled
mutually, namely, two Lorenz systems
\begin{equation}\label{eq:CoupledLorenzs}
\begin{array}{l}
\dot{x_{1,2}}=\sigma(y_{1,2}-x_{1,2})+\varepsilon(x_{2,1}-x_{1,2}),\\
\dot{y_{1,2}}=r_{1,2}x_{1,2}-y_{1,2}-x_{1,2}z_{1,2},\\
\dot{z_{1,2}}=-bz_{1,2}+x_{1,2}y_{1,2}.\\
\end{array}
\end{equation}
where $\mathbf{x}_{1,2}(t)=(x_{1,2},y_{1,2},z_{1,2})^T$ are the
vector-states of the interacting systems, $\varepsilon$ is a
coupling parameter, $\sigma=10.0$, $b=8/3$, $r_1=40.0$ and $r_2=35.0$. Due to the
bistable type of the chaotic attractor of the Lorenz oscillator
there are certain particularities of the GS regime which do not take
place in the systems with the R\"ossler-like chaotic attractor.

\begin{figure}[tb]
\centerline{\scalebox{0.4}{\includegraphics{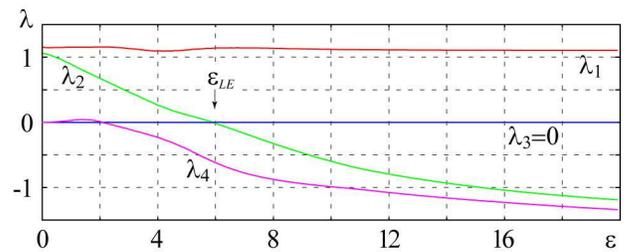}}}
\caption{(Color online) The dependencies of the four largest
Lyapunov exponents on the coupling strength $\varepsilon$}
\label{fgr:Lorenz}
\end{figure}

\begin{figure}[tb]
\centerline{\scalebox{0.4}{\includegraphics{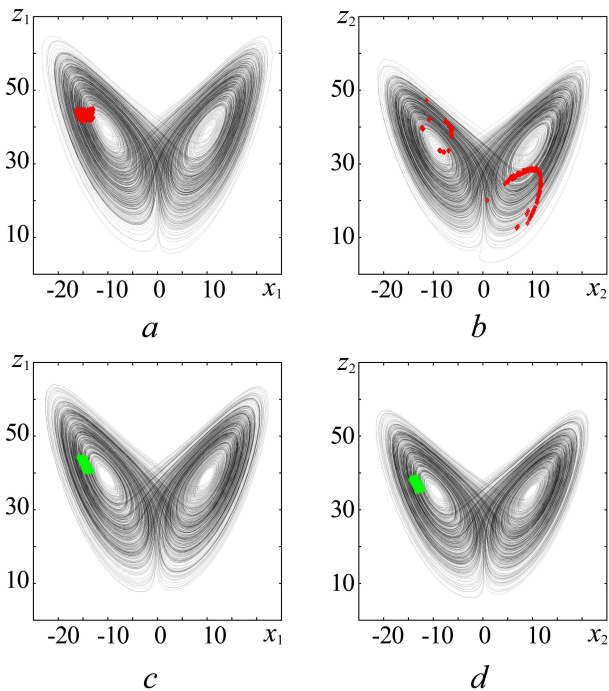}}}
\caption{(Color online) The phase portraits of two mutually coupled Lorenz
oscillators~(\ref{eq:CoupledLorenzs}) for $\varepsilon=5.7$
(\textit{a,b}) and $\varepsilon=6.1$ (\textit{c,d}).
Figures (\textit{a,c}) show the chaotic attractor of the first system $\mathbf{x}(t)$ with the reference point $\mathbf{x}^{k}$
and its nearest neighbors $\mathbf{x}^{kn}$. Figures~(\textit{b,d}) illustrate the corresponding
to them states $\mathbf{u}^{k,kn}$ in the phase space of the second system
$\mathbf{u}(t)$} \label{fgr:LorenzNN}
\end{figure}

According to the decision made in
Sec.~\ref{sct:LE&NNM}-\ref{sct:MutuallyCoupledPierceDiodes}, the
onset of the GS regime is connected with the sign change of the
initially positive Lyapunov exponent $\lambda_2$ at
$\varepsilon_{LE}\approx6.0$ (Fig.~\ref{fgr:Lorenz}). To prove this
statement for two mutually coupled Lorenz
systems~(\ref{eq:CoupledLorenzs}) we have used the nearest neighbor
method again both below (Fig.~\ref{fgr:LorenzNN},\,\textit{a,b}) and
above (Fig.~\ref{fgr:LorenzNN},\,\textit{c,d}) the critical point
$\varepsilon_{LE}$. For this purpose the reference point
$\mathbf{x}^{k}$ and its nearest neighbors $\mathbf{x}^{kn}$ have
been selected in the phase space of the first Lorenz system
(Fig.~\ref{fgr:LorenzNN},\,\textit{a,c}) and the corresponding to
them points $\mathbf{u}^{k,kn}$ have been found in the the phase
space of the second Lorenz oscillator
(Fig.~\ref{fgr:LorenzNN},\,\textit{b,d}). For
$\varepsilon>\varepsilon_{LE}$ the existence of the GS regime is
evidenced by the fact that all states of the second Lorenz system
corresponding to the nearest states of the first oscillator are also
nearest (Fig.~\ref{fgr:LorenzNN},\,\textit{d}). Alternatively, below
the threshold $\varepsilon$ the points of the second system are
located on the both sheets of the chaotic attractor
(Fig.~\ref{fgr:LorenzNN},\,\textit{b}) that indicates the break of
GS.

Again, as well as for two mutually coupled R\"ossler
systems~(\ref{eq:Roesslers}) \lulu{and Pierce diodes
(\ref{eq:PierceDiodes})}, the onset of the GS regime is shown to be
connected with the sign change of the initially positive Lyapunov
exponent $\lambda_2$ taking place when the coupling strength
$\varepsilon$ grows.

\begin{figure}[tb]
\centerline{\scalebox{0.4}{\includegraphics{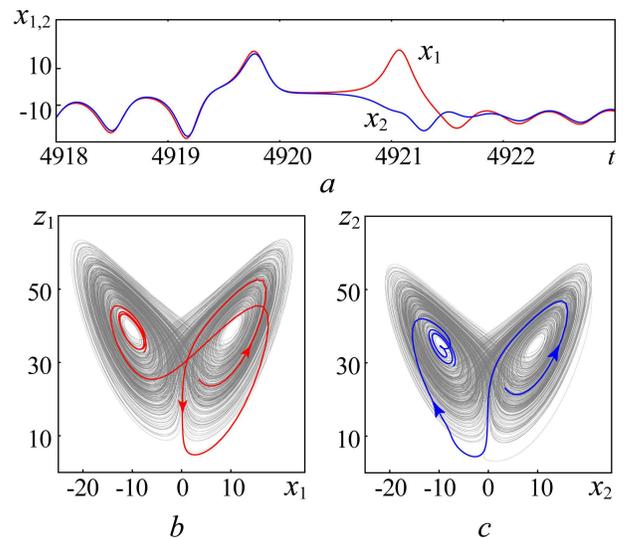}}}
\caption{(Color online) (\textit{a}) The fragment of the time series of two
mutually coupled Lorenz oscillators~(\ref{eq:CoupledLorenzs})
corresponding to the short-term period of the phase trajectory
divergence. (\textit{b}) The phase portraits of Lorenz systems and
the phase trajectories corresponding to the time interval shown in
Fig.~\ref{fgr:Lorenz_attr_5.7},\,\textit{a}. The coupling strength
$\varepsilon=5.7$, the GS regime is not observed} \label{fgr:Lorenz_attr_5.7}
\end{figure}

\begin{figure}[tb]
\centerline{\scalebox{0.4}{\includegraphics{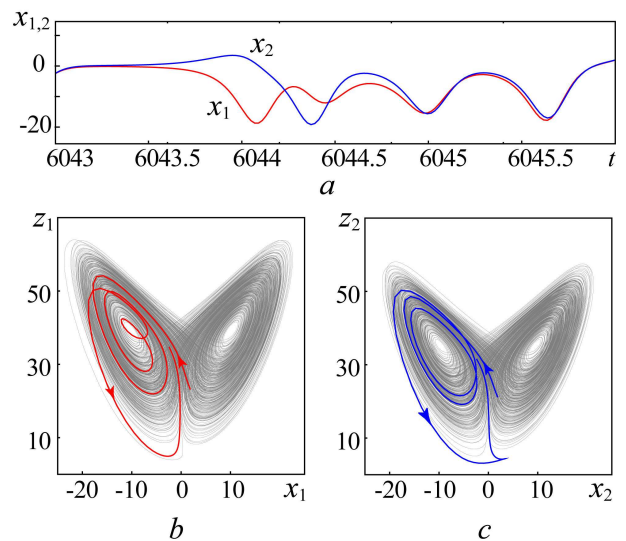}}}
\caption{(Color online) (\textit{a}) The fragment of the time series of two
mutually coupled Lorenz oscillators~(\ref{eq:CoupledLorenzs})
corresponding to the short-term period of the phase trajectory
divergence. (\textit{b}) The phase portraits of Lorenz systems and
the phase trajectories corresponding to the time interval shown in
Fig.~\ref{fgr:Lorenz_attr_6.1},\,\textit{a} The coupling strength
$\varepsilon=6.1$, the GS regime is detected} \label{fgr:Lorenz_attr_6.1}
\end{figure}

At the same time, in the dynamics of mutually coupled Lorenz
oscillators~(\ref{eq:CoupledLorenzs}) there are certain
particularities connected with the occurrence of GS caused by the
bistable type of the chaotic attractor of the system under
study~\footnote{E.g., this feature requires the proper choice of the
reference point for the nearest neighbor method.}. One can see
easily that in the case under consideration the coupling strength
value corresponding to the onset of GS is larger sufficiently than
the analogous value for two mutually coupled R\"ossler
systems~(\ref{eq:Roesslers}) \lulu{and Pierce diodes
(\ref{eq:PierceDiodes})}. Owing to the great coupling strength
value, in the vicinity of the GS onset (below the bifurcation point
$\varepsilon_{LE}$) two interacting Lorenz systems are greatly
synchronized with each other almost all time except for the short
time intervals when the representation point of one of the coupled
oscillators remains in the one sheet of the chaotic attractor
whereas the representation point of the second oscillator jumps to
another sheet (see~Fig.~\ref{fgr:Lorenz_attr_5.7}). After such a
short jump both phase trajectories approach each other, and the
oscillators start showing synchronous dynamics again. It is the
short-time phase trajectory divergence that is responsible for the
GS regime does not come into being. Above the critical point
$\varepsilon_{LE}$ there are also the phase trajectory divergencies,
but they do not envelop both sheets of attractors and the
representation points remain in the limits of one and the same sheet
during this perturbations (Fig.~\ref{fgr:Lorenz_attr_6.1}).

The difference between these types of dynamics observed below and
above the critical point $\varepsilon_{LE}$ allows to explain the
occurrence of the GS regime in two mutually coupled Lorenz
systems~(\ref{eq:CoupledLorenzs}). When the representation points
are located at one and the same sheet of the chaotic attractor, the
functional relation between vector states~(\ref{eq:MutualFunctRel})
is likely to exist, since the observed types of behavior is very
close to the complete synchronization regime due to the
rather large value of the coupling strength $\varepsilon$.
Alternatively, the divergence of the phase trajectories on the
different sheets of the chaotic attractor terminates this functional
relation, and, in turn, the generalized synchronization regime
is destroyed. So, below the critical point $\varepsilon_{LE}$ the GS
regime does not take place due to the presence of the short-term
time intervals with the divergence of the phase trajectories on the
different chaotic attractor sheets. More specifically, the
intermittent behavior near the onset of the GS regime is observed
(as well as in the vicinity of the other types of chaotic
synchronization like lag
synchronization~\cite{Boccaletti:2000_IntermitLagSynchro,Zhan:2002_Synchro},
phase
synchronization~\cite{Boccaletti:2002_LaserPSTransition_PRL,Hramov:RingIntermittency_PRL_2006}
and for generalized synchronization in the unidirectionally coupled
oscillators~\cite{Hramov:2005_IGS_EuroPhysicsLetters}), that may be
considered as the additional evidence of the correctness of the
obtained results. Note, that above the onset of the GS regime
($\varepsilon_{LE}$) there are also the phase trajectory
divergencies (which do not destroy the GS regime, since the
representation points remain in the limits of one and the same
attractor sheet and, as a consequence, the functional
relation~(\ref{eq:MutualFunctRel}) takes place) preventing the
occurrence of LS. These perturbations of the synchronous dynamics
vanish above the critical point $\varepsilon_{LS}$ where the lag
synchronization regime comes into being.

So, having considered the behavior of two mutually coupled Lorenz
system, we have obtained the additional evidence of the correctness
of the proposed viewpoint on the GS regime in the systems with the
mutual type of the coupling.

\lulu{\section{GS in networks of coupled
oscillators}\label{sct:Networks}}

Now we move to a more complicated situation, and analyze the GS in
complex networks. As we have mentioned above, in this case between
the interacting system states the functional relation in the form
(\ref{eq:NetworkFunctRel}) should be established.

Developing the concept of GS in the mutually coupled oscillators for
the networks, one can say that the phenomenon of GS in the complex
network can be understood as the state of the whole network when the
co-ordinates of all oscillators consisting this network are uniquely
determined by the values of co-ordinates of only one node
$\mathbf{x}_k$ (chosen arbitrary).
Following the arguments given in Sec.~\ref{sct:LE&NNM}, one can use
for the generalized synchronization regime the implicit form of the
functional relation between network's node states
\begin{equation}\label{eq:ImplicitFormForNetwork}
\mathbf{x}_i(t^*)=\mathbf{\tilde{F}}[\mathbf{x}_k(t^*)],\quad \forall i\neq k,
\end{equation}
where ${t^*-\delta<t<t^*+\delta}$ (where $\delta$ is infinitely
small). Again, locally, in the given range
$t^*-\delta<t<t^*+\delta$, we deal with the already known case.
Under assumption $N_d=3$ made above the following local Lyapunov
exponents characterize the dynamics of the systems: $\lambda^k_1>0$,
$\lambda^k_2=0$, $\lambda^k_3<0$, $\lambda^i_{1,2,3}<0$ ${\forall
i\neq k}$. In other words, in the selected area of the $3N$D
phase-space the manifold corresponding to the GS regime is
characterized by one unstable direction $\mathbf{e}^u$ and one
direction with the neutral stability $\mathbf{e}^0$ lying inside
this manifold, whereas all other directions are stable. These
directions correspond to one positive, one zero and $(3N-2)$
negative Lyapunov exponents. The same statement is also correct for
the other moments of time $t^*$, therefore, for the complex network
the manifold corresponding to the GS regime at every moment of time
is characterized by one unstable direction, one direction with the
neutral stability and $(3N-2)$ stable directions. So, having
calculated the spectrum of Lyapunov exponents for the network of
coupled chaotic oscillators one obtain that the GS regime in this
case produces one positive, one zero and $(3N-2)$ negative Lyapunov
exponents.

To prove the theoretical assumptions mentioned above we consider a
network consisting of $N=5$ R\"ossler systems with slightly
mismatched $\omega$-parameter values. The evolution of $i$-th
node(${i=1,\dots,N}$) is described by the following equations
\begin{equation}\label{eq:RosslerNetwork}
\begin{array}{ll}
\dot x_{i}=-\omega_{i}y_{i}-z_{i} +\varepsilon\sum_{j=1}^N G_{ij}x_j,\\
\dot y_{i}=\omega_{i}x_{i}+ay_{i},\\
\dot z_{i}=p+z_{i}(x_{i}-c),
\end{array}
\end{equation}
where the values of the control parameters $a$, $p$, $c$ have been
chosen to be the same of the case of two coupled
oscillators~(\ref{eq:Roesslers}), ${\omega_1=0.95}$,
${\omega_2=0.9525}$, ${\omega_3=0.955}$, ${\omega_4=0.9575}$,
${\omega_5=0.96}$, $\mathbf{x}_{i}(t)=(x_{i},y_{i},z_{i})^T$ is the
vector-state of the $i$-th node, $\varepsilon$ is the coupling
strength between nodes, $G_{ij}$ is the element of the coupling
matrix $\mathbf{G}$. $\mathbf{G}$ is a symmetric zero row sum
matrix, with $G_{ij}$ ($i\neq j$) being equal to 1 whenever node $i$
is connected with node $j$ and 0 otherwise, and $G_{ii}=-\sum_{j\neq
i}G_{ij}$. The topology of the links between nodes in the network
under study has been selected in such a way that each element of the
network is connected with each other.

\begin{figure}[tb]
\centerline{\scalebox{0.4}{\includegraphics{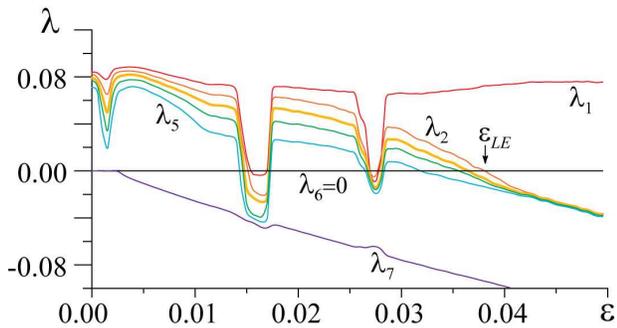}}}
\caption{(Color online) The dependencies of the seven largest Lyapunov exponents on
the coupling strength $\varepsilon$ for the network consisting of
five R\"ossler systems~(\ref{eq:RosslerNetwork}). The onset of the
GS regime in the network $\varepsilon_{LE}$ is shown by the arrow}
\label{fgr:NetworkLEs}
\end{figure}

\begin{figure*}[t]
\centerline{\scalebox{0.8}{\includegraphics{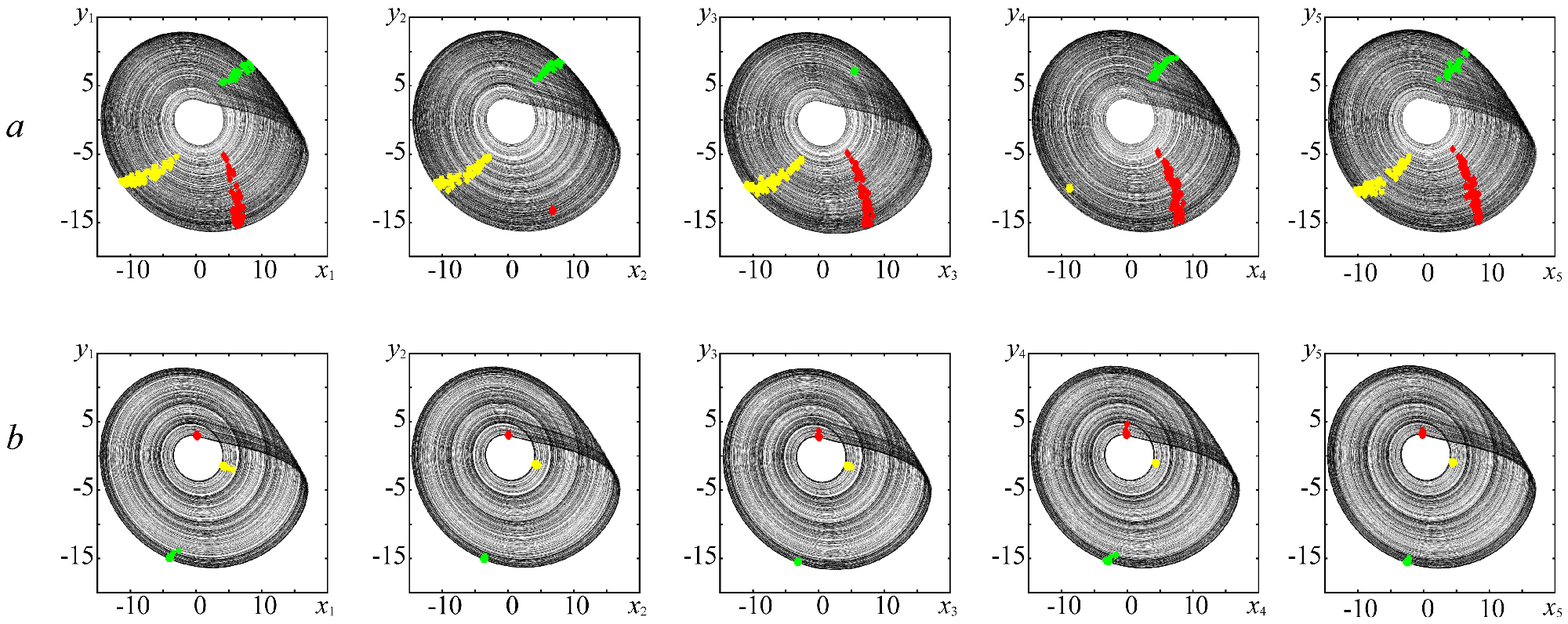}}}
\caption{(Color online) The phase portraits of five R\"ossler oscillators for two
different values of the coupling parameter: (\textit{a})
$\varepsilon=0.03$ (the PS regime) and (\textit{b})
$\varepsilon=0.04$ (the GS regime)} \label{fgr:RoesNetwNN}
\end{figure*}

The dynamics of the considered network is characterized by $3N=15$
Lyapunov exponents. If the coupling between nodes is equal to zero,
there are $N$ positive, $N$ negative, and $N$ zero Lyapunov
exponents. With the increase of the coupling strength $\varepsilon$
the zero Lyapunov exponents as well as the positive ones go
gradually to the region of the negative values. The dependencies of
the seven largest Lyapunov exponents on the coupling strength
$\varepsilon$ for the network consisting of five R\"ossler systems
are shown in Fig.~\ref{fgr:NetworkLEs}. One can see that at
$\varepsilon_{LE}\approx0.0385$ the second Lyapunov exponent
$\lambda_2$ passes through zero and becomes negative. Therefore, the
generalized synchronization regime is expected to be observed above
the critical value $\varepsilon_{LE}$.

To prove the presence of GS we have used the nearest neighbor method
in the same way as for two mutually coupled R\"ossler systems. In
Fig.~\ref{fgr:RoesNetwNN} the phase portraits of all R\"ossler
systems of the network are shown for two values of the coupling
strength, below (Fig.~\ref{fgr:RoesNetwNN},\,\textit{a},
$\varepsilon=0.03$) and above
(Fig.~\ref{fgr:RoesNetwNN},\,\textit{b}, $\varepsilon=0.04$) the
critical point $\varepsilon_{LE}$.

In the phase portraits of the three systems $\mathbf{x}_{i}(t)$,
$i=2\div4$ three points (one point for each system) with its nearest
neighbors have been selected randomly
(\textcolor{red}{$\blacklozenge$} --- $i=2$,
\textcolor{green}{$\mathbf{+}$} --- $i=3$,
\textcolor{yellow}{$\mathbf{\boxdot}$} --- $i=4$) and the points
corresponding to them have been detected in all other coupled
systems. For $\varepsilon=0.03$
(Fig.~\ref{fgr:RoesNetwNN},\,\textit{a}) the points are concentrated
in a limited range of attractor and distributed along the radius
being the evidence of the presence of PS and the absence of GS. For
$\varepsilon>\varepsilon_{LE}$
(Fig.~\ref{fgr:RoesNetwNN},\,\textit{b}) all states of all
oscillators are nearest neighbors, thus proving the existence of GS.

\begin{figure}[tb]
\includegraphics[width=8.2cm]{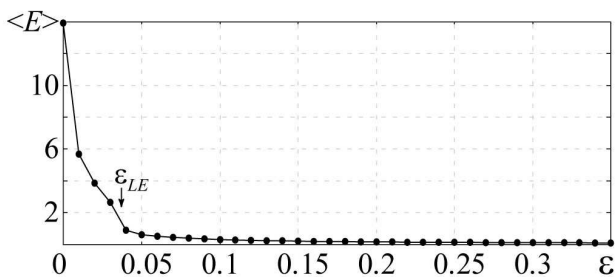}
\caption{$\langle E \rangle$ vs $\varepsilon$
for the network of R\"ossler oscillators  (\ref{eq:RosslerNetwork}).
The onset of the
GS regime in the network $\varepsilon_{LE}$ is shown by the arrow} \label{fgr:Erro}
\end{figure}

Similarly to the case of two mutually coupled R\"ossler oscillators
considered above we compare the onset of the GS and LS in the
network (\ref{eq:RosslerNetwork}). But due to the small values of
the control parameter detuning the LS regime is very close to the
complete synchronization (CS) one. In the simulations, the onset of
the CS regime can be monitored by looking at the vanishing of the
time average  (over a window $T$) synchronization error
\begin{equation}
 \langle E \rangle =
\frac{1}{T(N-1)}\sum_{j>1}\int^{t+T}_t\|\mathbf{x}_j-\mathbf{x}_1\|dt'.
\label{eq:Error}
\end{equation}
In the present case, we adopt as vector norm
$\|\mathbf{x}\|=\sqrt{x^2+y^2+z^2}$. Fig.~\ref{fgr:Erro} reports the
synchronization error versus $\varepsilon$ for a given topology. One
can see that the synchronization error becomes close to zero
considerably later the GS regime arising. When the GS regime takes
place $\langle E\rangle$ is still positive that is the evidence of
the CS regime absence.

\begin{figure}[t]
\includegraphics[width=8.2cm]{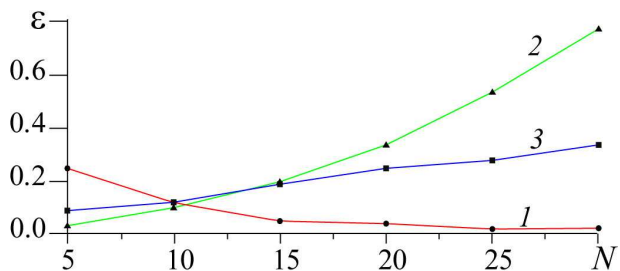}
\caption{(Color online) Boundaries of the GS regime on the ``number of elements $N$ --- coupling parameter $\varepsilon$''--plane for random (curve \textit{1}), regular (curve \textit{2}) and ``small world'' (curve \textit{3}) networks} \label{fgr:LargeNet}
\end{figure}

So, the GS regime in networks of coupled nonlinear elements can be
detected by the moment of transition of the second (positive)
Lyapunov exponent in the field of the negative values. Now we
analyze the influence of the number of elements and topology of the
network on the GS regime onset. In Fig.~\ref{fgr:LargeNet} the
boundaries of the GS regime on the ``number of elements $N$ ---
coupling parameter $\varepsilon$''--plane for networks of different
topologies of links between nodes are shown. Curve \textit{1}
corresponds to the random network whereas curves \textit{2} and
\textit{3} refer to the regular and ``small-world'' networks,
respectively. For all considered cases the values of the control
parameters $\omega_i$ have been selected randomly in such a way that
the probability distribution density of $\omega_i$-values has been
obeyed by the Gaussian distribution with the mean value
$\omega_0=0.95$ and variance $\Delta\omega=0.017$ that corresponds
to the case of the relatively large values of the control parameter
detuning. It is clearly seen from Fig.~\ref{fgr:LargeNet} that the
topology of the network influences sufficiently on the GS regime
onset. In particular, the threshold of the GS regime onset decreases
for the random network, whereas both for the regular and ``small
world'' ones it increases monotonically. At that, the boundary of GS
for regular network grows more rapidly in comparison with the
``small world'' one.

\section{Mechanisms of GS occurrence}
\label{sct:Mechanisms}

Now, following the approach of some of our previous
works~\cite{Aeh:2005_GS:ModifiedSystem}, we move to reveal the
mechanisms associated with the emergence of the GS regime in the
case of a generic ensemble of \lulu{coupled} systems (with,
obviously, the case of two coupled oscillators considered as the
simplest variant of such a configuration).

For the purpose of exemplification, and without lack of generality,
we here-below characterize the state of the network by the only
vector ${\mathbf{U}=(u_1,u_2,\dots,u_i,\dots,u_{N\cdot N_d})^T}$,
where $u_{3i-2}=x_i$, $u_{3i-1}=y_i$, $u_{3i}=z_i$, instead of the
set of vectors ${\mathbf{x}_i=(x_i,y_i,z_i)^T}$, $i=\overline{1,N}$.
The dimension of each element of the network is assumed to be
$N_d=3$ again, but this analytical study may be extended easily to
the other systems with arbitrary dimensions $N_d$.

Following the above formalism, the entire network may be considered
as a high-dimensional autonomous dynamical system, whose evolution
equation is given by
\begin{equation}\label{eq:NetworkEquation}
\mathbf{\dot U}=\mathbf{L}(\mathbf{U})+\varepsilon\mathbf{\tilde{G}}\mathbf{U}.
\end{equation}
Here the vector function $\mathbf{L}(\cdot)$ determines the
evolution of the elements of the network in the absence of the
coupling, whereas the additive term $\varepsilon\mathbf{\tilde{G}}$
describes the influence of the topology and the coupling strength of
the links between oscillators. Matrix $\mathbf{\tilde{G}}$ specifies
the structure of the dissipative couplings between nodes, and it is
assumed to be a symmetric zero row sum matrix,
$\tilde{G}_{ii}=-\sum_{j\neq i}\tilde{G}_{ij}$, with
$\tilde{G}_{ij}$ ($i\neq j$) being equal to 1 whenever variable
$u_i$ forces the variable $u_j$ and 0 otherwise.

It is easy to see that the term
$\varepsilon\mathbf{\tilde{G}}\mathbf{U}$ brings the additional
dissipation into the system~(\ref{eq:NetworkEquation}). Indeed, the
phase flow contraction is characterized by means of the vector field
divergence
\begin{equation}
\lim\limits_{\Delta t\rightarrow 0}\lim\limits_{\Delta V\rightarrow 0}\frac{1}{\Delta V}\frac{\Delta V}{\Delta t}=
\mathrm{div}\,\mathbf{L}+\varepsilon\sum\limits_{i=1}^{N_dN}\tilde{G}_{ii},
\end{equation}
where $\Delta V$ is the elementary volume of the phase space of the
system~(\ref{eq:NetworkEquation}). Since $\tilde{G}_{ii}\leq 0$, the
term $\varepsilon\sum_{i=1}^{N_dN}\tilde{G}_{ii}$ is also negative
and the dissipation in the considered \lulu{group} increases with
the growth of the coupling strength $\varepsilon$, resulting in the
simplification of the otherwise chaotic dynamics of the
system~(\ref{eq:NetworkEquation}).

To characterize the complexity of the  motion, the spectrum of
Lyapunov exponents  is frequently used. In the case under study,
let's suppose that the behavior of the
system~(\ref{eq:NetworkEquation}) is initially described by the set
${\lambda_1\geq\lambda_2\geq\dots\geq\lambda_{NN_d}}$ of Lyapunov
exponents, with $N$ of them being positive (i.e., while no
synchronous motion is set up, each one of the elements of the
network contributes independently with one positive Lyapunov
exponent to the overall spectrum). As  dissipation increases, some
of the initially positive Lyapunov exponents  become negative, and
each passage of a Lyapunov exponent trough zero testifies that one
more degree of freedom of the chaotic motion corresponds to a
contractive direction. When $\lambda_2$ becomes negative, only one
degree of freedom is representative of the evolution of the network,
i.e. a GS regime is built. Indeed, as soon as $\lambda_2$ is
negative, all systems have to arrange their evolution into a
specific collective motion, wherein the functional
relation~(\ref{eq:NetworkFunctRel}) is taking place.

\section*{CONCLUSION}
\label{sct:Conclusion}

In conclusion, we have analyzed the GS regime in systems with a
mutual type of the coupling. We have extended the definition of GS,
for being valid also for pairs of mutually coupled chaotic
oscillators, and for the complex networks. The GS regime onset
in 3D systems is shown to be connected with the
zero-crossing of the second Lyapunov exponent. GS may be, therefore,
considered in terms of the transition from the high-dimensional
hyperchaotic regime to the chaotic oscillations. The obtained
results are proved by means of the nearest neighbor method.  Since
the developed theory is applicable to different systems, we expect
that the very same mechanism will be observed in many other relevant
circumstances.
Particularly, the obtained results could be extended to
the systems which dimension of the phase space $N_d>3$ including the
spatially extended media and coupled systems with a different
dimension of the phase space.



\begin{thebibliography}{34}
\expandafter\ifx\csname
natexlab\endcsname\relax\def\natexlab#1{#1}\fi
\expandafter\ifx\csname bibnamefont\endcsname\relax
  \def\bibnamefont#1{#1}\fi
\expandafter\ifx\csname bibfnamefont\endcsname\relax
  \def\bibfnamefont#1{#1}\fi
\expandafter\ifx\csname citenamefont\endcsname\relax
  \def\citenamefont#1{#1}\fi
\expandafter\ifx\csname url\endcsname\relax
  \def\url#1{\texttt{#1}}\fi
\expandafter\ifx\csname urlprefix\endcsname\relax\def\urlprefix{URL
}\fi \providecommand{\bibinfo}[2]{#2}
\providecommand{\eprint}[2][]{\url{#2}}

\bibitem[{\citenamefont{Boccaletti
  et~al.}(2002{\natexlab{a}})\citenamefont{Boccaletti, Kurths, Osipov,
  Valladares, and Zhou}}]{Boccaletti:2002_ChaosSynchro}
\bibinfo{author}{\bibfnamefont{S.}~\bibnamefont{Boccaletti}},
  \bibinfo{author}{\bibfnamefont{J.}~\bibnamefont{Kurths}},
  \bibinfo{author}{\bibfnamefont{G.~V.} \bibnamefont{Osipov}},
  \bibinfo{author}{\bibfnamefont{D.~L.} \bibnamefont{Valladares}},
  \bibnamefont{and} \bibinfo{author}{\bibfnamefont{C.~S.} \bibnamefont{Zhou}},
  \bibinfo{journal}{Physics Reports} \textbf{\bibinfo{volume}{366}},
  \bibinfo{pages}{1} (\bibinfo{year}{2002}{\natexlab{a}}).

\bibitem[{\citenamefont{Abarbanel et~al.}(1996)\citenamefont{Abarbanel, Rulkov,
  and Sushchik}}]{Rulkov:1996_AuxiliarySystem}
\bibinfo{author}{\bibfnamefont{H.~D.~I.} \bibnamefont{Abarbanel}},
  \bibinfo{author}{\bibfnamefont{N.~F.} \bibnamefont{Rulkov}},
  \bibnamefont{and} \bibinfo{author}{\bibfnamefont{M.~M.}
  \bibnamefont{Sushchik}}, \bibinfo{journal}{Phys. Rev. E}
  \textbf{\bibinfo{volume}{53}}, \bibinfo{pages}{4528} (\bibinfo{year}{1996}).

\bibitem[{\citenamefont{Hramov and
  Koronovskii}(2005{\natexlab{a}})}]{Aeh:2005_GS:ModifiedSystem}
\bibinfo{author}{\bibfnamefont{A.~E.} \bibnamefont{Hramov}} \bibnamefont{and}
  \bibinfo{author}{\bibfnamefont{A.~A.} \bibnamefont{Koronovskii}},
  \bibinfo{journal}{Phys. Rev. E} \textbf{\bibinfo{volume}{71}},
  \bibinfo{pages}{067201} (\bibinfo{year}{2005}{\natexlab{a}}).

\bibitem[{\citenamefont{Kocarev and Parlitz}(1996)}]{Kocarev:1996_GS}
\bibinfo{author}{\bibfnamefont{L.}~\bibnamefont{Kocarev}} \bibnamefont{and}
  \bibinfo{author}{\bibfnamefont{U.}~\bibnamefont{Parlitz}},
  \bibinfo{journal}{Phys. Rev. Lett.} \textbf{\bibinfo{volume}{76}},
  \bibinfo{pages}{1816} (\bibinfo{year}{1996}).

\bibitem[{\citenamefont{Zheng and Hu}(2000)}]{Zhigang:2000_GSversusPS}
\bibinfo{author}{\bibfnamefont{Z.}~\bibnamefont{Zheng}} \bibnamefont{and}
  \bibinfo{author}{\bibfnamefont{G.}~\bibnamefont{Hu}}, \bibinfo{journal}{Phys.
  Rev. E} \textbf{\bibinfo{volume}{62}}, \bibinfo{pages}{7882}
  (\bibinfo{year}{2000}).

\bibitem[{\citenamefont{Hramov et~al.}(2005{\natexlab{a}})\citenamefont{Hramov,
  Koronovskii, and Popov}}]{Hramov:2005_GLEsPRE}
\bibinfo{author}{\bibfnamefont{A.~E.} \bibnamefont{Hramov}},
  \bibinfo{author}{\bibfnamefont{A.~A.} \bibnamefont{Koronovskii}},
  \bibnamefont{and} \bibinfo{author}{\bibfnamefont{P.~V.} \bibnamefont{Popov}},
  \bibinfo{journal}{Phys. Rev. E} \textbf{\bibinfo{volume}{72}},
  \bibinfo{pages}{037201} (\bibinfo{year}{2005}{\natexlab{a}}).

\bibitem[{\citenamefont{Rulkov}(1996)}]{Rulkov:1996_SynchroCircuits}
\bibinfo{author}{\bibfnamefont{N.~F.} \bibnamefont{Rulkov}},
  \bibinfo{journal}{Chaos} \textbf{\bibinfo{volume}{6}}, \bibinfo{pages}{262}
  (\bibinfo{year}{1996}).

\bibitem[{\citenamefont{Rogers et~al.}(2004)\citenamefont{Rogers, Kalra,
  Schroll, Uchida, Lathrop, and Roy}}]{GS_LightModulator}
\bibinfo{author}{\bibfnamefont{E.~A.} \bibnamefont{Rogers}},
  \bibinfo{author}{\bibfnamefont{R.}~\bibnamefont{Kalra}},
  \bibinfo{author}{\bibfnamefont{R.~D.} \bibnamefont{Schroll}},
  \bibinfo{author}{\bibfnamefont{A.}~\bibnamefont{Uchida}},
  \bibinfo{author}{\bibfnamefont{D.~P.} \bibnamefont{Lathrop}},
  \bibnamefont{and} \bibinfo{author}{\bibfnamefont{R.}~\bibnamefont{Roy}},
  \bibinfo{journal}{Phys.Rev.Lett.} \textbf{\bibinfo{volume}{93}},
  \bibinfo{pages}{084101} (\bibinfo{year}{2004}).

\bibitem[{\citenamefont{Dmitriev et~al.}(2009)\citenamefont{Dmitriev, Hramov,
  Koronovskii, Starodubov, Trubetskov, and Zharkov}}]{dmitriev:074101}
\bibinfo{author}{\bibfnamefont{B.~S.} \bibnamefont{Dmitriev}},
  \bibinfo{author}{\bibfnamefont{A.~E.} \bibnamefont{Hramov}},
  \bibinfo{author}{\bibfnamefont{A.~A.} \bibnamefont{Koronovskii}},
  \bibinfo{author}{\bibfnamefont{A.~V.} \bibnamefont{Starodubov}},
  \bibinfo{author}{\bibfnamefont{D.~I.} \bibnamefont{Trubetskov}},
  \bibnamefont{and} \bibinfo{author}{\bibfnamefont{Y.~D.}
  \bibnamefont{Zharkov}}, \bibinfo{journal}{Physical Review Letters}
  \textbf{\bibinfo{volume}{102}}, \bibinfo{pages}{074101}
  (\bibinfo{year}{2009}).

\bibitem[{\citenamefont{Hramov et~al.}(2008{\natexlab{a}})\citenamefont{Hramov,
  Koronovskii, and Popov}}]{Hramov:2008_INIS_PRE}
\bibinfo{author}{\bibfnamefont{A.~E.} \bibnamefont{Hramov}},
  \bibinfo{author}{\bibfnamefont{A.~A.} \bibnamefont{Koronovskii}},
  \bibnamefont{and} \bibinfo{author}{\bibfnamefont{P.~V.} \bibnamefont{Popov}},
  \bibinfo{journal}{Phys. Rev. E} \textbf{\bibinfo{volume}{77}},
  \bibinfo{pages}{036215} (\bibinfo{year}{2008}{\natexlab{a}}).

\bibitem[{\citenamefont{Terry and VanWiggeren}(2001)}]{Terry:GSchaosCom2001}
\bibinfo{author}{\bibfnamefont{J.}~\bibnamefont{Terry}} \bibnamefont{and}
  \bibinfo{author}{\bibfnamefont{G.}~\bibnamefont{VanWiggeren}},
  \bibinfo{journal}{Chaos, Solitons and Fractals}
  \textbf{\bibinfo{volume}{12}}, \bibinfo{pages}{145} (\bibinfo{year}{2001}).

\bibitem[{\citenamefont{Koronovskii et~al.}(2009)\citenamefont{Koronovskii,
  Moskalenko, and Hramov}}]{alkor:2010_SecureCommunicationUFNeng}
\bibinfo{author}{\bibfnamefont{A.~A.} \bibnamefont{Koronovskii}},
  \bibinfo{author}{\bibfnamefont{O.~I.} \bibnamefont{Moskalenko}},
  \bibnamefont{and} \bibinfo{author}{\bibfnamefont{A.~E.}
  \bibnamefont{Hramov}}, \bibinfo{journal}{Physics-Uspekhi}
  \textbf{\bibinfo{volume}{52}}, \bibinfo{pages}{1213} (\bibinfo{year}{2009}).

\bibitem[{\citenamefont{Moskalenko et~al.}(2010)\citenamefont{Moskalenko,
  Koronovskii, and Hramov}}]{Moskalenko:InfoTransNoisePLA2010}
\bibinfo{author}{\bibfnamefont{O.~I.} \bibnamefont{Moskalenko}},
  \bibinfo{author}{\bibfnamefont{A.~A.} \bibnamefont{Koronovskii}},
  \bibnamefont{and} \bibinfo{author}{\bibfnamefont{A.~E.}
  \bibnamefont{Hramov}}, \bibinfo{journal}{Phys. Lett. A}
  \textbf{\bibinfo{volume}{374}}, \bibinfo{pages}{2925} (\bibinfo{year}{2010}).

\bibitem[{\citenamefont{Rulkov et~al.}(1995)\citenamefont{Rulkov, Sushchik,
  Tsimring, and Abarbanel}}]{Rulkov:1995_GeneralSynchro}
\bibinfo{author}{\bibfnamefont{N.~F.} \bibnamefont{Rulkov}},
  \bibinfo{author}{\bibfnamefont{M.~M.} \bibnamefont{Sushchik}},
  \bibinfo{author}{\bibfnamefont{L.~S.} \bibnamefont{Tsimring}},
  \bibnamefont{and} \bibinfo{author}{\bibfnamefont{H.~D.~I.}
  \bibnamefont{Abarbanel}}, \bibinfo{journal}{Phys. Rev. E}
  \textbf{\bibinfo{volume}{51}}, \bibinfo{pages}{980} (\bibinfo{year}{1995}).

\bibitem[{\citenamefont{Pecora and Carroll}(1991)}]{Pecora:1991_ChaosSynchro}
\bibinfo{author}{\bibfnamefont{L.~M.} \bibnamefont{Pecora}} \bibnamefont{and}
  \bibinfo{author}{\bibfnamefont{T.~L.} \bibnamefont{Carroll}},
  \bibinfo{journal}{Phys. Rev. A} \textbf{\bibinfo{volume}{44}},
  \bibinfo{pages}{2374} (\bibinfo{year}{1991}).

\bibitem[{\citenamefont{Pyragas}(1997)}]{Pyragas:1997_CLEsFromTimeSeries}
\bibinfo{author}{\bibfnamefont{K.}~\bibnamefont{Pyragas}},
  \bibinfo{journal}{Phys. Rev. E} \textbf{\bibinfo{volume}{56}},
  \bibinfo{pages}{5183} (\bibinfo{year}{1997}).

\bibitem[{\citenamefont{Pyragas}(1996)}]{Pyragas:1996_WeakAndStrongSynchro}
\bibinfo{author}{\bibfnamefont{K.}~\bibnamefont{Pyragas}},
  \bibinfo{journal}{Phys. Rev. E} \textbf{\bibinfo{volume}{54}},
  \bibinfo{pages}{R4508} (\bibinfo{year}{1996}).

\bibitem[{Spr(2002)}]{Springer:2002_EncyclopaediaofMathematics}
\emph{\bibinfo{title}{Encyclopaedia of Mathematics}}
  (\bibinfo{publisher}{Springer-Verlag Berlin Heidelberg New York},
  \bibinfo{year}{2002}), \bibinfo{edition}{{M}ichiel {H}azewinkel} ed.

\bibitem[{\citenamefont{Parlitz et~al.}(1996)\citenamefont{Parlitz, Junge,
  Lauterborn, and Kocarev}}]{Parlitz:1996_PhaseSynchroExperimental}
\bibinfo{author}{\bibfnamefont{U.}~\bibnamefont{Parlitz}},
  \bibinfo{author}{\bibfnamefont{L.}~\bibnamefont{Junge}},
  \bibinfo{author}{\bibfnamefont{W.}~\bibnamefont{Lauterborn}},
  \bibnamefont{and} \bibinfo{author}{\bibfnamefont{L.}~\bibnamefont{Kocarev}},
  \bibinfo{journal}{Phys. Rev. E} \textbf{\bibinfo{volume}{54}},
  \bibinfo{pages}{2115} (\bibinfo{year}{1996}).

\bibitem[{\citenamefont{Hramov et~al.}(2005{\natexlab{b}})\citenamefont{Hramov,
  Koronovskii, and Moskalenko}}]{Harmov:2005_GSOnset_EPL}
\bibinfo{author}{\bibfnamefont{A.~E.} \bibnamefont{Hramov}},
  \bibinfo{author}{\bibfnamefont{A.~A.} \bibnamefont{Koronovskii}},
  \bibnamefont{and} \bibinfo{author}{\bibfnamefont{O.~I.}
  \bibnamefont{Moskalenko}}, \bibinfo{journal}{Europhysics Letters}
  \textbf{\bibinfo{volume}{72}}, \bibinfo{pages}{901}
  (\bibinfo{year}{2005}{\natexlab{b}}).

\bibitem[{\citenamefont{Hramov et~al.}(2007)\citenamefont{Hramov, Koronovskii,
  and Kurovskaya}}]{Hramov:2007_2TypesPSDestruction}
\bibinfo{author}{\bibfnamefont{A.~E.} \bibnamefont{Hramov}},
  \bibinfo{author}{\bibfnamefont{A.~A.} \bibnamefont{Koronovskii}},
  \bibnamefont{and} \bibinfo{author}{\bibfnamefont{M.~K.}
  \bibnamefont{Kurovskaya}}, \bibinfo{journal}{Phys. Rev. E}
  \textbf{\bibinfo{volume}{75}}, \bibinfo{pages}{036205}
  (\bibinfo{year}{2007}).

\bibitem[{\citenamefont{Hramov et~al.}(2008{\natexlab{b}})\citenamefont{Hramov,
  Koronovskii, and Kurovskaya}}]{Hramov:ZeroLE_PRE2008}
\bibinfo{author}{\bibfnamefont{A.~E.} \bibnamefont{Hramov}},
  \bibinfo{author}{\bibfnamefont{A.~A.} \bibnamefont{Koronovskii}},
  \bibnamefont{and} \bibinfo{author}{\bibfnamefont{M.~K.}
  \bibnamefont{Kurovskaya}}, \bibinfo{journal}{Phys. Rev. E}
  \textbf{\bibinfo{volume}{78}}, \bibinfo{pages}{036212}
  (\bibinfo{year}{2008}{\natexlab{b}}).

\bibitem[{\citenamefont{Godfrey}(1987)}]{Godfrey:1987}
\bibinfo{author}{\bibfnamefont{B.~B.} \bibnamefont{Godfrey}},
  \bibinfo{journal}{Phys. Fluids} \textbf{\bibinfo{volume}{30}},
  \bibinfo{pages}{1553} (\bibinfo{year}{1987}).

\bibitem[{\citenamefont{Matsumoto et~al.}(1996)\citenamefont{Matsumoto,
  Yokoyama, and Summers}}]{Matsumoto:1996}
\bibinfo{author}{\bibfnamefont{H.}~\bibnamefont{Matsumoto}},
  \bibinfo{author}{\bibfnamefont{H.}~\bibnamefont{Yokoyama}}, \bibnamefont{and}
  \bibinfo{author}{\bibfnamefont{D.}~\bibnamefont{Summers}},
  \bibinfo{journal}{Phys.Plasmas} \textbf{\bibinfo{volume}{3}},
  \bibinfo{pages}{177} (\bibinfo{year}{1996}).

\bibitem[{\citenamefont{Trubetskov and
  Hramov}(2003)}]{TrueAeh:2003_2004_microwave_electronics_1_2engl}
\bibinfo{author}{\bibfnamefont{D.~I.} \bibnamefont{Trubetskov}}
  \bibnamefont{and} \bibinfo{author}{\bibfnamefont{A.~E.}
  \bibnamefont{Hramov}}, \emph{\bibinfo{title}{Lectures on microwave
  electronics for physicists}} (\bibinfo{publisher}{Vol.~1,2. Fizmatlit,
  Moscow}, \bibinfo{year}{2003}).

\bibitem[{\citenamefont{Filatov et~al.}(2006)\citenamefont{Filatov, Hramov, and
  Koronovskii}}]{Filatov:2006_PierceDiode_PLA}
\bibinfo{author}{\bibfnamefont{R.~A.} \bibnamefont{Filatov}},
  \bibinfo{author}{\bibfnamefont{A.~E.} \bibnamefont{Hramov}},
  \bibnamefont{and} \bibinfo{author}{\bibfnamefont{A.~A.}
  \bibnamefont{Koronovskii}}, \bibinfo{journal}{Phys. Lett. A}
  \textbf{\bibinfo{volume}{358}}, \bibinfo{pages}{301} (\bibinfo{year}{2006}).

\bibitem[{\citenamefont{Filatova et~al.}(2008)\citenamefont{Filatova, Hramov,
  Koronovskii, and Boccaletti}}]{filatova:023133}
\bibinfo{author}{\bibfnamefont{A.}~\bibnamefont{Filatova}},
  \bibinfo{author}{\bibfnamefont{A.}~\bibnamefont{Hramov}},
  \bibinfo{author}{\bibfnamefont{A.}~\bibnamefont{Koronovskii}},
  \bibnamefont{and}
  \bibinfo{author}{\bibfnamefont{S.}~\bibnamefont{Boccaletti}},
  \bibinfo{journal}{Chaos: An Interdisciplinary Journal of Nonlinear Science}
  \textbf{\bibinfo{volume}{18}}, \bibinfo{eid}{023133}
  (pages~\bibinfo{numpages}{6}) (\bibinfo{year}{2008}).

\bibitem[{\citenamefont{Rouch}(1976)}]{Rouch:1976_FluidNumericalBook}
\bibinfo{author}{\bibfnamefont{P.~J.} \bibnamefont{Rouch}},
  \emph{\bibinfo{title}{Computational fluid dynamics}}
  (\bibinfo{publisher}{Hermosa publishers, Albuquerque}, \bibinfo{year}{1976}).

\bibitem[{\citenamefont{Hramov et~al.}(2012)\citenamefont{Hramov, Koronovskii,
  Maksimenko, and Moskalenko}}]{hramov:SLE_POP2012}
\bibinfo{author}{\bibfnamefont{A.~E.} \bibnamefont{Hramov}},
  \bibinfo{author}{\bibfnamefont{A.~A.} \bibnamefont{Koronovskii}},
  \bibinfo{author}{\bibfnamefont{V.~A.} \bibnamefont{Maksimenko}},
  \bibnamefont{and} \bibinfo{author}{\bibfnamefont{O.~I.}
  \bibnamefont{Moskalenko}}, \bibinfo{journal}{Physics of Plasmas}
  \textbf{\bibinfo{volume}{19}}, \bibinfo{pages}{082302} (\bibinfo{year}{2012}).

\bibitem[{\citenamefont{Boccaletti and
  Valladares}(2000)}]{Boccaletti:2000_IntermitLagSynchro}
\bibinfo{author}{\bibfnamefont{S.}~\bibnamefont{Boccaletti}} \bibnamefont{and}
  \bibinfo{author}{\bibfnamefont{D.~L.} \bibnamefont{Valladares}},
  \bibinfo{journal}{Phys. Rev. E} \textbf{\bibinfo{volume}{62}},
  \bibinfo{pages}{7497} (\bibinfo{year}{2000}).

\bibitem[{\citenamefont{Zhan et~al.}(2002)\citenamefont{Zhan, Wei, and
  Lai}}]{Zhan:2002_Synchro}
\bibinfo{author}{\bibfnamefont{M.}~\bibnamefont{Zhan}},
  \bibinfo{author}{\bibfnamefont{G.~W.} \bibnamefont{Wei}}, \bibnamefont{and}
  \bibinfo{author}{\bibfnamefont{C.~H.} \bibnamefont{Lai}},
  \bibinfo{journal}{Phys. Rev. E} \textbf{\bibinfo{volume}{65}},
  \bibinfo{pages}{036202} (\bibinfo{year}{2002}).

\bibitem[{\citenamefont{Boccaletti
  et~al.}(2002{\natexlab{b}})\citenamefont{Boccaletti, Allaria, Meucci, and
  Arecchi}}]{Boccaletti:2002_LaserPSTransition_PRL}
\bibinfo{author}{\bibfnamefont{S.}~\bibnamefont{Boccaletti}},
  \bibinfo{author}{\bibfnamefont{E.}~\bibnamefont{Allaria}},
  \bibinfo{author}{\bibfnamefont{R.}~\bibnamefont{Meucci}}, \bibnamefont{and}
  \bibinfo{author}{\bibfnamefont{F.~T.} \bibnamefont{Arecchi}},
  \bibinfo{journal}{Phys. Rev. Lett.} \textbf{\bibinfo{volume}{89}},
  \bibinfo{pages}{194101} (\bibinfo{year}{2002}{\natexlab{b}}).

\bibitem[{\citenamefont{Hramov et~al.}(2006)\citenamefont{Hramov, Koronovskii,
  Kurovskaya, and Boccaletti}}]{Hramov:RingIntermittency_PRL_2006}
\bibinfo{author}{\bibfnamefont{A.~E.} \bibnamefont{Hramov}},
  \bibinfo{author}{\bibfnamefont{A.~A.} \bibnamefont{Koronovskii}},
  \bibinfo{author}{\bibfnamefont{M.~K.} \bibnamefont{Kurovskaya}},
  \bibnamefont{and}
  \bibinfo{author}{\bibfnamefont{S.}~\bibnamefont{Boccaletti}},
  \bibinfo{journal}{Phys. Rev. Lett.} \textbf{\bibinfo{volume}{97}},
  \bibinfo{pages}{114101} (\bibinfo{year}{2006}).

\bibitem[{\citenamefont{Hramov and
  Koronovskii}(2005{\natexlab{b}})}]{Hramov:2005_IGS_EuroPhysicsLetters}
\bibinfo{author}{\bibfnamefont{A.~E.} \bibnamefont{Hramov}} \bibnamefont{and}
  \bibinfo{author}{\bibfnamefont{A.~A.} \bibnamefont{Koronovskii}},
  \bibinfo{journal}{Europhysics Lett.} \textbf{\bibinfo{volume}{70}},
  \bibinfo{pages}{169} (\bibinfo{year}{2005}{\natexlab{b}}).

\end{thebibliography}

\end{document}